\documentclass[11pt, letter]{article}

\usepackage{graphicx,amsmath,amsfonts,multirow, natbib, authblk, url}
\usepackage[margin=1in]{geometry}
\newcommand{\HC}{{\tt HyDA}}
\newcommand{\Real}{\mathbb{R}}

\begin{document}
%\firstpage{1}

\title{\bf Distilled Single Cell Genome Sequencing and \emph{De Novo} Assembly for Sparse Microbial Communities}
\author[1]{Zeinab Taghavi\footnote{To whom correspondence should be addressed.}\thanks{ztaghavi@wayne.edu}}
\author[1]{Narjes S. Movahedi\thanks{narges@wayne.edu}}
\author[1,2]{Sorin Draghici\thanks{sorin@wayne.edu}}
\author[1]{Hamidreza Chitsaz\thanks{chitsaz@wayne.edu}}
\affil[1]{Department of Computer Science, Wayne State University, Detroit, MI 48202}
\affil[2]{Department of Obstetrics and Gynecology, Wayne State University, Detroit, MI 48202}

\maketitle

\begin{abstract}

Identification of every single genome present in a microbial sample is an important and challenging task with crucial applications. It is challenging because there are typically millions of cells  in a microbial sample, the vast majority of which elude cultivation. The most accurate method to date is exhaustive single cell sequencing using multiple displacement amplification, which is simply intractable for a large number of cells. However, there is hope for breaking this barrier as the number of different cell types with distinct genome sequences is usually much smaller than the number of cells.

Here, we present a novel divide and conquer method to sequence and \emph{de novo} assemble all distinct genomes present in a microbial sample with a sequencing cost and computational complexity proportional to the number of genome types, rather than the number of cells. The method is implemented in a tool called Squeezambler. We evaluated Squeezambler on simulated data. The proposed divide and conquer method successfully reduces the cost of sequencing in comparison with the na\"ive exhaustive approach.

{\bf Availability:} Squeezambler and datasets are available under \\
{\tt http://compbio.cs.wayne.edu/software/squeezambler/}

\end{abstract}

\section{Introduction}
Critical applications, including the Human Microbiome Project \citep{Methe12}, biothreat detection, and combating antibiotic resistant pathogens, necessitate identification of all distinct genome sequences in a bacterial sample. When prior knowledge is available about which organisms may be present in the sample, flow cytometry and 16S rRNA gene sequencing may be used. However, metagenomics is usually used for analyzing the genomics of relatively abundant microbes when no prior knowledge is given. Metagenomics consists in the study of the variation of species in a complex microbial sample. Since the vast majority of environmental bacteria elude cultivation, metagenomics investigates microbial communities by sequencing sampled genome fragments without the need for culturing. Such a heterogeneous pool of sequencing reads can also be assembled to yield a superposition of highly abundant genomes in the sample \citep{Treangen13}. There are two problems with metagenomics: (i) the resulting assembly contains multiple genomes superimposed, and (ii) only highly abundant genomes survive the sampling process. 

\begin{figure}[!tpb]
\begin{center}
\vspace*{0.2in}
\includegraphics[width=0.6\textwidth]{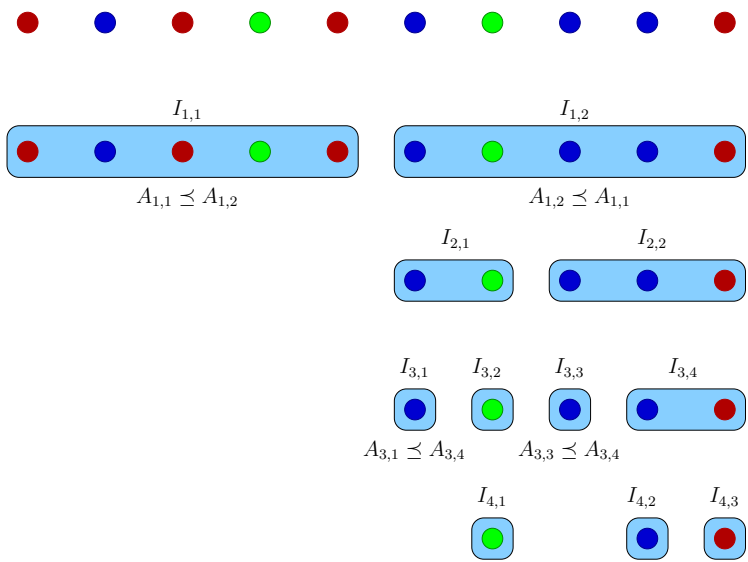}
\end{center}
\caption{The divide and conquer algorithm for an example with 10 cells and 3 distinct genomes shown in different colors. Each row corresponds to one sequencing round. The number of barcodes in each round is the number of blue boxes in the corresponding row. }\label{fig:example}
\end{figure}

Advances in DNA amplification technology have enabled whole genome sequencing directly from individual cells without requiring growth in culture. Single cell sequencing methods have enabled investigation of novel uncultured microbes \citep{Kvist07,Mussmann07}. These culture-independent single cell studies are a powerful alternative to metagenomics studies. Genomic sequencing from single bacterial genomes was first demonstrated with cells isolated by flow cytometry \citep{Raghunathan05}, using multiple displacement amplification (MDA) \citep{Dean02,Dean01,Hosono03}. MDA is now the preferred method for whole genome amplification from single cells \citep{Lasken07,Ishoey08}. The first attempt to assemble a complete bacterial genome from one cell \citep{Zhang06} further explored the challenges of assembly from amplified DNA, including amplification bias and chimeric DNA rearrangements. Amplification bias results in orders of magnitude difference in coverage \citep{Raghunathan05}, and absence of coverage in some regions. Chimera formation occurs during the DNA branching process by which the $\phi_{29}$ DNA polymerase generates DNA amplification in MDA \citep{Lasken07a}. Subsequent studies continued to improve single cell assemblies \citep{Marcy07,Podar07,Hongoh08,Rodrigue09,Woyke09}. A nearly full potential of single cell genome assembly has recently been realized by the work of \citealp{Chitsaz11} followed by those of \citealp{Bankevich12} and \citealp{Peng12}.

Due to recent progress in single cell DNA amplification techniques and \emph{de novo} assembly algorithms, the genomes of all bacterial species in a sample can be captured one cell at a time. However, there are often millions of cells per sample, in which case the na\"ive deep sequencing of every cell becomes prohibitively costly. Moreover, it is expected that deep sequencing of every cell is often not necessary as the majority of biologically important samples are sparse in the sense that many cells are biological replicates. Compressed \citep{CanTao05,CanTao06,Donoho06} and distilled (adaptive sampling-and-refinement) sensing methods \citep{Haupt11,WeiHer12} have been proposed in the last decade to exploit sparsity for reducing the cost of sensing and reconstructing signals in various spaces, ranging from Banach spaces to Boolean algebras \citep{Erlich10,Stobbe12}. Inspired by those advances, we give an algorithm in this paper that exploits sample sparsity to reduce the cost of sequencing without compromising the accuracy of identification of all distinct genomes, even the ones that are minimally represented in the sample.

\section{Approach}
A na\"ive approach to solve the problem, which we call {\it single cell co-assembly strategy}, is to amplify the genome of each cell, barcode them individually, pool them, sequence in one sequencing run, and demultiplex based on the barcode. In this approach, each cell should be isolated and its DNA extracted and amplified using multiple displacement amplification. Although there is currently no high throughput device to perform these processes, one could envision automated microfluidic devices that will be capable of high throughput separation, DNA extraction, amplification, and barcoding of single cells in the near future. The output sequencing reads could  then be co-assembled using our tool {\tt HyDA} \citep{Movahedi12}. In {\tt HyDA}, the read dataset of each cell is assigned a unique color. All the colors are co-assembled in one colored de Bruijn graph. This approach requires enough unique barcodes to tag the fragments of each cell. Also, barcodes attached to each fragment are sequenced, which imposes additional sequencing cost. Fabrication of so many unique barcodes becomes prohibitively expensive for a large number of cells and therefore, this    na\"ive approach  will not work.

The number of distinct genomes in a microbial community is often much less than the number of cells. For example, \citealp{Qin10} estimated the number of detected distinct species in the human gut to be in the order of 1,000, while the number of microbial cells in a human body, most of which reside in the gut, is in the order of 100 trillion. We call this effect the {\it sparsity} of distinct genomes in a sizable microbial population. The na\"ive approach does not exploit the sparsity to reduce the cost of sequencing. Here, we proposed to exploit the spareness by  adopting a { \it divide and conquer strategy} to reduce the amount of required barcodes and sequencing. After isolation of each cell and extraction of the DNA, every DNA is amplified and stored separately, e.g. in a microfluidic droplet. The main idea is to sample the amplified DNAs adaptively, which is essentially allocating sequencing and barcoding resources dynamically over the course of multiple sensing iterations. Initially, the algorithm has one group of cells, which is the entire sample. In each iteration, each group is divided into two equally-sized subgroups. A small amount of DNA from each cell in a subgroup is sampled, pooled, and sequenced. In practice, one barcode per subgroup is used to multiplex and demultiplex the sequencing in one run. The amount of sampling from each cell is computed based on the results of previous iterations. This is called resource allocation and is similar to the one proposed by \citealp{Haupt11} and \citealp{WeiHer12}. The resulting read datasets, one per subgroup, are then co-assembled and compared using {\tt HyDA} to decide pairwise subsumption of subgroups. The cells in those subgroups that are subsumed by other subgroups, even in previous iterations, are eliminated from further analysis. This procedure is continued until each remaining group contains only one cell. The resulting non-redundant single element groups capture all of the distinct genome sequences. 

%\begin{methods}
%\section{Preliminaries}

\section{Methods}
Although distinct genomes are often identified as different species, there are numerous cases where distinct genomes are categorized as varied instances of the same species or even the same strain. Instead of identification of strains and species which are currently phenotypic notions, the goal of our approach is to find all distinct genomes in a sample. We define two genome sequences to be distinct if the ratio of their differences over the whole genome size is above a threshold. That threshold is input by the user and controls a trade-off between sensitivity and specificity. 

Co-assembly and comparison of multiple input read datasets lie at the core of both approaches we take in this paper. While there are assembly tools for single cell genomic data, such as {\tt SPAdes} \citep{Bankevich12} and {\tt IDBA-UD} \citep{Peng12}, and also co-assembly tools for normal multicell genomic data such as {\tt Cortex} \citep{Iqbal12}, we use {\tt HyDA} which is the only tool to date that has both functionalities \citep{Movahedi12}. 
However, the novel ideas proposed here can also be implemented using other assemblers. For the sake of completeness, {\tt HyDA} algorithm is summarized in the following. 

\subsection{Co-assembly Algorithm}

\subsubsection{Construction and Condensation of the Colored De Bruijn Graph}\label{sec:hyda-graph}
The colored de Bruijn graph, a variation of the standard de Bruijn graph, is a combinatorial structure that can be used to assemble a number of input read datasets, each represented by a color, superimposed on a single de Bruin graph \citep{Iqbal12}. The output of such assembly methods is a number of assembled sequences (contigs) and the corresponding average multiplicity in each color. Our de Bruijn graph of the input reads is stored in a hashed collection of splay trees whose vertices are $k$-mers with an array of multiplicity counts (one entry per color), in- and out-edges, and internal flags. A 1-in 1-out chain of $k$-mers is condensed into an equivalent long sequence which is called a \emph{unitig}. A maximal unitig, which cannot be extended further due to a branch in the graph, is a \emph{contig}. Note that in {\tt HyDA}, our condensation is solely based on the topology of the graph without any attention to the colored multiplicities. Ignoring multiplicities for condensation is purposefully done, and constitutes the  feature that allows the assembler to deal with black out regions in single cell multiple displacement amplification \citep{Chitsaz11}. 

Contigs with low coverage are often caused by sequencing error. The low coverage contig removal process is iterated with an increased cutoff in each round. In each iteration, those contigs whose maximum coverage over all colors is less than the cutoff are eliminated, and the remaining graph is recondensed. This causes some contigs to merge into larger ones with recomputed average coverages. This process is similar to {\tt Velvet-SC}'s low coverage contig removal, but instead of considering one average coverage per contig, \HC\ considers  the \emph{maximum} of average coverages for all the colors  of each contig \citep{Chitsaz11}. In this case, only those contigs that have low coverage in all colors are considered erroneous and removed. Another possible approach is to eliminate those contigs for which the mean of average coverages for all colors is less than the cutoff. However, if we were to follow this approach, a contig that is well covered in one color but is poorly covered or absent in hundreds of colors would be lost since the mean would dilute the effect of coverage in one color among hundreds of colors.  This approach would not work for us here  because our goal is to be able to preserve rare contigs.

\subsubsection{Redundancy Removal}
To remove redundant genomes, we define a relation that is reminiscent of subset relation on the set of contigs for each color. Note that our goal here is to remove redundant genomes, which are collections of contigs, rather than to remove individually redundant contigs. Let $A = \{a_1, a_2, \ldots, a_r\}$ be the set of remaining contigs after iterative error removal. Let $M_j(a_i)$ denote the average coverage of contig $a_i$ in color $j$, for $1 \leq i \leq r$ and $1 \leq j \leq s$, where $s$ is the number of colors. Pick $\epsilon \geq 0$ and let $A_j = \{a_i \in A\ | \ M_j(a_i) > \epsilon\} \subset A$ be the set of contigs for each color $j$. The parameter $\epsilon$ determines the trade-off between specificity and sensitivity. We chose $\epsilon = 0$ in this study, but a non-zero $\epsilon$ might be needed if there are erroneous or contaminant $k$-mers in one color that also occur in the true genomic sequence of another color. 

We define $D_\tau(A_i,A_j) \in \Real$ on the set ${\mathcal F} = \{ A_1, A_2, \ldots, A_s\}$ as:
\begin{align} \label{equ:D}
D_\tau(A_i,A_j) = \tau - \frac{\|A_{i} \backslash A_{j}\|}{\|A_{i}\|},
\end{align}
in which $A_{i} \backslash A_{j} = \{a \in A_i | a \notin A_j \}$, and $\|\cdot\|$ denotes the total assembly size. We define:
\begin{align} \label{equ:relation}
A_i \preceq_\tau A_j \mbox{ iff } 0 \leq D_\tau(A_i,A_j), 
\end{align}
in which $\tau \geq 0$. Particularly, $\preceq_0$ becomes the subset relation and can be used to detect and remove redundant collections of contigs, i.e. those that are subsumed by a larger collection. However in reality, the mathematical subset relation is not adequate as there are various types of noise including sequencing errors, intrastrain variations such as SNPs and indels, contaminations added in the amplification and sequencing process, and lack of coverage in some regions caused by the MDA. Hence, the definition of subset should be loosened by choosing a small but nonzero value for $\tau$. Beside $\epsilon$, the value of $\tau$ gives a trade-off between specificity and sensitivity of recognizing distinct genomes. If $\tau$ is small, the algorithm detects two equivalent genomes as distinct, and if $\tau$ is large, distinct genomes are declared equivalent. To see how $\tau$ is chosen, refer to Sections \ref{sec:dividandconquer} and \ref{sec:simulationresults}. The results are shown in Table \ref{tab:results}. Finally, we compute a non-redundant set of assemblies ${\mathcal E} = \{ A_{i_1}, A_{i_2}, \ldots, A_{i_t}\} \subseteq {\mathcal F}$ such that for every distinct pair $1 \leq a,b \leq t$, $A_{i_a} \not \preceq_\tau A_{i_b}$ and $A_{i_b} \not \preceq_\tau A_{i_a}$.

%\subsubsection{Maximal Contig Sets As Distinct Comprehensive Genomic Sequences.}
%Maximal elements in ${\mathcal F}$ are distinct comprehensive genomic sequences, e.g. if each color represents one single cell, then such maximal elements are the assemblies of candidate species. They are candidate and not precisely different species, because it is possible to have two maximal contig sets for the same species due to lack of coverage in some regions caused by MDA. As the coverage of each cell increases, the possibility of having distinct maximal sets for one species decreases. 

%Pritchard has given a sub-quadratic algorithm to compute the maximal elements in a partial order induced by subset relation \citep{Pritchard99}. It is practically more efficient than the na\"ive all-pairs comparison if $s$, the number of colors, is large. For this study, we implemented the na\"ive algorithm to compute the maximal elements of $\mathcal F$. We plan to implement the Pritchard's algorithm in the future.
\subsection{Divide and Conquer Strategy Exploiting Sparsity \label{sec:dividandconquer}}

Let $n$ be the number of cells in the sample, and denote the cells by $S^i, i = 1, \ldots, n$. Our algorithm aims to assemble all of the distinct genomes and identify at least one cell per distinct genome. To reach this goal, our algorithm iteratively pools samples of amplicons from different cells, tags each pool with a unique barcode, mixes the barcoded pools, and has the result sequenced. The objective is to  minimize the total number of bases required to be sequenced as well as the number of different barcodes needed. 

In the first iteration, we divide the $n$ cells $S^1, \ldots, S^n$ into two sets  $$I_{1,1}= \{S^1, \ldots, S^{\lfloor n/2 \rfloor} \},$$ $$I_{1,2}= \{S^{\lfloor n/2 \rfloor+1}, \ldots, S^n \}.$$ Our algorithm samples equal amount of amplicons from each cell in $I_{1,1}$ and $I_{1,2}$. The amplicons in each set are pooled and tagged by two distinct barcodes. The barcoded amplicons are sequenced to reach a desired number of base pairs. This number is an input parameter of our algorithm. We define the total number of base pairs sequenced from $I_{1,i}$ to be $b_{1,i}$, for $i=1, 2$. The two read datasets are co-assembled by {\tt HyDA} using two colors. The result is two sets of contigs for each color, $A_{1,1}$ and $A_{1,2}$. We calculate $D_{\tau_1}(A_{1,1},A_{1,2})$ and  $D_{\tau_1}(A_{1,2},A_{1,1})$ as defined in (\ref{equ:D}), in which $\tau_1=\tau / \max_j |I_{1,j}|$, $\tau$ is an input parameter, and $|\cdot|$ is the set cardinality. We choose $\tau$ to be the maximum allowable difference between the assembly of two single cells from the same strain. Based on these values, we decide if the relations $A_{1,1} \preceq_{\tau_1} A_{1,2}$ and $A_{1,2} \preceq_{\tau_1} A_{1,1}$ hold. If $A_{1,1}$ is a subset of $A_{1,2}$, then all of the distinct genomes in  $I_{1,1}$ are present in $I_{1,2}$, therefore, the cells in $I_{1,1}$ do not need further sampling. This applies to $I_{1,2}$ too, if $A_{1,2}$ is a subset of $A_{1,1}$. If both relations hold, one of $I_{1,1}$ or $I_{1,2}$ is eliminated arbitrarily from further analysis. Each remaining set $I_{1,\cdot}$ is divided into two subsets for analysis in the second iteration. Fig. \ref{fig:example} depicts an example of 10 cells with 3 distinct genomes shown in different colors.

The same splitting process occurs in the subsequent iterations. Assume $I_{i,1}, \ldots, I_{i,m_i}$ are the remaining sets in iteration $i$. Each set $I_{i,j}$ is sampled to produce $b_{i,j}$ base pairs, barcoded uniquely, pooled, and sequenced. All of the new sequence datasets and those obtained in all previous iterations are co-assembled. In the co-assembly, the previous datasets help to improve the assembly of the new ones. The resulting contig set of $I_{i,j}$ is denoted by $A_{i,j}$. For all $j,k =1 \ldots m_i$, the relations $A_{i,j} \preceq_{\tau_i} A_{i,k}$ are evaluated, where $\tau_i$ is a threshold whose calculation will be explained below. The cells in those $I_{i,j}$ whose assemblies are subsumed will be removed from further analysis. All the remaining ones are partitioned into two disjoint subsets. Denote the new subsets by $I_{i+1,1}, \ldots, I_{i+1,m_{i+1}}$. Note that in iteration $i$, $m_i$ unique barcodes are needed. Therefore, 
\begin{equation}\label{equ:m}
m=\max_i m_i 
\end{equation}
is the maximum number of barcodes required for the entire algorithm.

\begin{table*}[!tpb]
\begin{center}
\fontsize{9}{11}\selectfont 
\caption{The 9 chosen species for our simulation.\label{tab:species}}
{\begin{tabular}{llclr}\hline
NCBI ID & Name & Ref. Status & Size (bps) & No. of Cells \\\hline
NC\_004663.1 & Bacteroides thetaiotaomicron VPI-5482 chromosome & complete & 6.29 M & 23 \\
NC\_009614.1 & Bacteroides vulgatus ATCC 8482 chromosome & complete & 5.16 M & 7 \\
NC\_009615.1 & Parabacteroides distasonis ATCC 8503 chromosome & complete & 4.81 M & 8 \\
NC\_008532.1 & Streptococcus thermophilus LMD-9 & complete & 1.86 M & 2 \\
NC\_016776.1 & Bacteroides fragilis 638R & complete & 5.37 M & 1 \\\hline
FP929046.1 & Faecalibacterium prausnitzii SL3/3 & draft & 3.21 M & 12 \\
FP929051.1 & Ruminococcus bromii L2-63 & draft & 2.25 M & 35 \\
FP929053.1 & Ruminococcus sp. SR1/5 & draft & 3.55 M & 12 \\
FP929055.1 & Ruminococcus torques L2-14 & draft & 3.34 M  & 15 \\
\hline
\end{tabular}}{}
\end{center}
\end{table*}

Parameters $b_{i,j}$ play a key role in the algorithm. We propose an adaptive calculation of $b_{i,j}$ to minimize, without losing accuracy, the total base pairs sequenced: 
\begin{equation}\label{equ:b}
b = \sum_{i,j} b_{i,j}.
\end{equation}
Assume $I_{i,j}$ is a set that is created by dividing the set $I_{i-1,k}$ in iteration $i-1$ into two. We are motivated to choose the total number of sequenced base pairs from cells in $I_{i,j}$ to be proportional to the total assembly size $\|A_{i-1, k}\|$, i.e. 
\begin{equation} \label{equ:c}
b_{i,j} = c \times \|A_{i-1, k}\|,
\end{equation}
where $c$ is an input parameter indicating the estimated average coverage of each iteration. We say $A_{i,j}$ are \emph{accurate enough} if the partial order relation $\preceq_{\tau_i}$ can be assessed accurately. If $c$ is large and the assembly of iteration $i-1$ is accurate enough, then in iteration $i$, adequate base pairs are sequenced to allow an accurate enough assembly of set $I_{i,j}$. 

Another factor that affects accuracy of the assessment of these relations is the choice of $\tau_i$. The threshold $\tau_i$ in the $i^{\rm{th}}$ iteration is used to detect cells with similar genomes in spite of small differences in their assemblies. We propose to use the following threshold:
\begin{equation} \label{equ:tau_i}
\tau_i = \frac{\tau}{\max_{ 1 \leq j \leq m_i} |I_{i,j}|}.
\end{equation}
Recall that $\tau$ is the maximum allowable difference between the assembly of two single cells with similar genome sequences. To account for the worst possible case, it is assumed that there are $|I_{i,j}|$ distinct genomes in each group $I_{i,j}$. Therefore, $\max_{ 1 \leq j \leq m_i} |I_{i,j}| $ captures the maximum number of distinct genomes in $I_{i,j}$ from any $I_{i,k}$. With these assumptions, $\tau_i$ is a conservative threshold. This threshold will guarantee that two distinct genomes are detected, but it has the possibility of detecting similar genomes as distinct. In the last iteration of the algorithm, when every group consists of one cell, the threshold is $\tau$. Note that the number of iterations, which is the number of sequencing rounds, is always $\left\lceil \log_2 n \right\rceil$.
 %\textcolor{red}{It would be useful to have a big O estimate or such for the number of sequencing steps, not only the number of iterations. Note that in the example the sequencing is performed 11 times for only 10 cells and 3 species. In this example, it is more efficient to sequence every cell from the beginning. An estimate based on some reasonable assumptions will drive the point home that the number of times the sequencing is performed will track the number of species, not the number of cells.}

\subsection{Implementation}
We implemented our algorithm in a tool called {\tt Squeezambler 1.0} in C++. Our tool and datasets are available under   http://compbio.cs.wayne.edu/software/squeezambler/

%\end{methods}

\section{Results}
Since we did not have access to real data, we tested our algorithm using simulated data. We used our tool {\tt MDAsim 1.0}  \citep{Taghavi12} to simulate 100 multiple displacement amplification processes (one process per cell) from 9 distinct genomes. The output of {\tt MDAsim 1.0} was fed into an Illumina read generator, {\tt ART} \citep{Huang12}, to
generate Illumina reads, with realistic errors, from the simulated amplicons. The set of generated reads for each cell were treated like a microfluidic droplet from which samples without replacement are extracted in each iteration of {\tt Squeezambler 1.0}. We assume that MDA products are contamination free, which requires a contaminant free automated microfluidic cell sorting, amplification, and sampling device.

\subsection{Datasets}
Totally 115 MDAs were simulated from 9 distinct genomes chosen from the list of species found in a gut metagenomics study \citep{Qin10}, that have a complete or draft reference genome. Recall that we are simulating MDA from a reference genome, therefore, we needed a reference genome for the chosen species. Table \ref{tab:species} summarizes the NCBI ID, name, size, reference status (complete or draft), and the number of cells we have simulated. The number of cells is approximately proportional to the abundance mean of the corresponding species in \citep{Qin10}. We ran {\tt MDAsim 1.0} with a diverse set of parameters, one for each cell, to capture the diverse nature of MDA coverage bias. {\tt ART}, an Illumina read generator, was then deployed to generate 100 bps Illumina reads from the simulated amplicons. The amplification gain of {\tt MDAsim 1.0} was 300$\times$ and that of {\tt ART} was 8$\times$ from which $1/8$ were selected randomly to obtain a total gain of 300$\times$. We assembled the dataset of each cell individually with {\tt HyDA 1.0}, and the resulting assemblies have between 0.1\% and 4.2\% missing reference bases measured by {\tt Gage} \citep{Salzberg12}, which is similar to the real world situation with a successful MDA reaction \citep{Chitsaz11}. We made an error profile that matches the error statistics of Illumina HiSeq 2000 for {\tt ART}. Using our profile, {\tt ART} injects on average 1\% error into the reads, because of which we need to eliminate erroneous contigs in the assembler. {\tt HyDA 1.0}, and also its predecessor {\tt Velvet-SC}, have an iterative algorithm to remove low coverage contigs, which is explained in Section \ref{sec:hyda-graph}. In each iteration, {\tt Squeezambler 1.0} provides {\tt HyDA 1.0} with a coverage cutoff as a percentage of the mean coverage of each color. That percentage is constant in all iterations. 
 
We designed three collections of cells, the statistics of which are summarized in Table \ref{tab:experiments}. In the first collection, there are 62 cells with 6 distinct genomes. In this collection, we put 22 different MDA results from NC\_004663.1 and 22 from FP929051.1 to play the role of highly abundant genomes in a sample as well as 1 from NC\_016776.1 and 2 from NC\_008532.1 to represent low abundance genomes in the same sample. In the first collection, the number of distinct genomes is approximately one tenth of the number of cells but with a wide range of abundances. The second collection is the sparsest collection among the three, where the number of distinct genomes is approximately one twentieth of the number of cells. The third collection is the most diverse of the three, where the number of distinct genomes is approximately one sixteenth of the number of cells. 

\begin{table}[h!]
\begin{center}
\fontsize{9}{11}\selectfont 

\caption{Our simulation setups: (i) 62 cells; 6 species, (ii) 97 cells; 5 species, and (iii) 112 cells; 7 species.\label{tab:experiments}}
{\begin{tabular}{lcccccc}\hline
\multirow{3}{*}{NCBI ID} & \multicolumn{6}{c}{Abundance (\%)} \\
 & \multicolumn{2}{c}{62 cells; 6 species} & \multicolumn{2}{c}{97 cells; 5 species} & \multicolumn{2}{c}{112 cells; 7 species} \\\hline
NC\_004663.1 & 22 & 36\% & 23 & 24\% & 23 & 21\%\\
NC\_009614.1 & 7 & 11\% & 0 & 0\% & 7 & 6\%\\
NC\_009615.1 & 8 & 13\% & 0 & 0\% & 8 & 7\%\\
NC\_008532.1 & 2 & 3\% & 0 & 0\% & 0 & 0\%\\
NC\_016776.1 & 1 & 1\% & 0 & 0\% & 0 & 0\%\\
FP929046.1 & 0 & 0\% & 12 & 12\% & 12 & 11\%\\
FP929051.1 & 22 & 36\% & 35 & 36\% & 35 & 31\%\\
FP929053.1 & 0 & 0\% & 12 & 12\% & 12 & 11\%\\
FP929055.1 & 0 & 0\% & 15 & 16\% & 15 & 13\%\\
\hline
\end{tabular}}{}
\end{center}
\end{table}

\subsection{Simulation Results \label{sec:simulationresults}}

\begin{table}[h!]
%\begin{center}
\fontsize{9}{11}\selectfont 
\caption{{\tt Squeezambler 1.0} results for the three setups summarized in Table \ref{tab:experiments}. For some methods we report the results for two different values of initial sequencing coverage per cell.\label{tab:results}}
{\begin{tabular}{l|lccccc}\hline
Setup  & Method & Sequencing & Total & Max & No. of & Iterations\\
 & & per Cell$^*$ in the & Sequencing$^{**}$ & Barcodes$^{***}$ & Predicted & \\
  &  & $1^{\rm{st}}$ Iteration  & (bps) & &  Distinct &\\
  &  & (bps) & & &  Genomes &\\\hline
62 cells; & single cell co-assembly & 63 M & 3.9 G & 62 & 6 & 1\\
6 distinct & divide and conquer & 7 M & {\bf 3.0 G} & {\bf 10} & 8 & 6\\
genomes & & & & & & \\\hline
97 cells; & single cell co-assembly & 63 M & 6.1 G & 97 & 5 & 1\\
5 distinct & single cell co-assembly & 31 M & 3.0 G & 97 & 11 & 1\\
genomes & divide and conquer & 1 M & 3.2 G & 17 & 5 & 7 \\
& divide and conquer & 7 M & {\bf 2.9 G} & {\bf 10} & 6 & 7\\\hline
112 cells; & single cell co-assembly & 63 M & 7.1 G & 112 & 7 & 1\\
7 distinct & single cell co-assembly & 31 M & {\bf 3.5 G} & 112 & 14 & 1 \\
genomes & divide and conquer & 1 M & 7.1 G & {\bf 33} & 11 & 7\\
\hline
\end{tabular}}{
$^* b_{1,j} / |I_{1,j}|$.\\
$^{**} b$ in (\ref{equ:b}).\\
$^{***} m$ in (\ref{equ:m}).}
\end{table}

We ran {\tt Squeezambler 1.0} for the three collections described above, the results of which are summarized in Table \ref{tab:results}. Most of the {\tt Squeezambler 1.0} parameters were the same for all three collections. The assembly inclusion threshold constant per cell was chosen $\tau = 0.2$ which means at most 20\% of the assembly can vary among multiple instances of the same genome. Taking into account the genomic sequence loss caused by the MDA, sampling of the amplicons, and sequencing errors, 20\% is a reasonable choice. This is not an optimized value and is chosen based on the authors' intuition.  We chose $\tau$ conservatively in this work so that distinct genomes are not lost but some equivalent genomes are detected as distinct. Finding the optimal value for $\tau$ needs a thorough study which is beyond the scope of this paper.

The $k$-mer size for {\tt HyDA 1.0} was chosen to be $k = 55$ which is a common choice for the chosen Illumina error profile \citep{Chitsaz11}. The coverage cutoff, as a percentage of the coverage mean, was chosen to be 100\%, and the minimum contig length was 100 bps for {\tt HyDA 1.0}. The coverage mean is estimated based on the assembly size in the first iteration, which is often larger than the actual genome size due to a myriad of erroneous low coverage $k$-mers. This causes the initially estimated coverage mean to be a small fraction of the final coverage mean after error removal. 

{\tt Squeezambler 1.0} has an option to choose the number of initial groups in the first iteration, $g$. If $g$ is chosen to be equal to the number of cells, then {\tt Squeezambler 1.0} simulates the single cell co-assembly of all the cells. If $g = 2$, then {\tt Squeezambler 1.0} simulates the divide and conquer algorithm described in Section \ref{sec:dividandconquer}. Although experimenting with different $g$ values may improve the results, we do not have data for it. 

Before any sequencing is done, the algorithm has no idea about the genome sizes, various distinct genomes, and abundance of each genome. Therefore, an unbiased algorithm has to sequence from each cell exactly the same amount right in the $1^{\rm{st}}$ iteration. That amount in our algorithm, denoted by $b_{1,j} / |I_{1,j}|$, is an input parameter to {\tt Squeezambler 1.0} which was chosen to be between 1 Mbps and 63 Mbps as reported in the third column of Table \ref{tab:results}. In our setup, the size of the 9 distinct genomes varies between 1.8 Mbps and 6.3 Mbps; see Table \ref{tab:species}. Therefore, 1 Mbps sequencing provides between 1/6$\times$ and 1/2$\times$ coverage, and 63 Mbps sequencing provides between 10$\times$ and 35$\times$ coverage. 

The input parameter $c$, which controls the amount of sequencing in subsequent rounds, was chosen to be $c = 10$, which means 10$\times$ expected coverage from each genome in each collection. We observed that in practice $10\times$ coverage of each distinct genome provides sufficient information for reliable evaluation of $\preceq$. This is consistent with the \citealp{Lander88} analysis, in which the statistics of gaps and contigs in terms of coverage is characterized. Based on that analysis, $10\times$ coverage yields the entire genome without gap with high probability.

Our divide and conquer algorithm exhibits significant improvement in maximum barcodes, and in most cases the total number of base pairs sequenced, over the single cell co-assembly method. For the 97 cells, 5 distinct genomes collection, our divide and conquer algorithm requires only 2.9 Gbps sequencing and 10 barcodes in comparison to 3.0 Gbps sequencing and 97 barcodes consumed by the single cell co-assembly method. Similarly for the 62 cells, 6 distinct genomes collection, our divide and conquer algorithm requires only 3.0 Gbps sequencing and 10 barcodes in comparison to 3.9 Gbps sequencing and 62 barcodes required by the single cell co-assembly method. Even though for the 112 cells, 7 distinct genomes collection, our divide and conquer algorithm outperforms single cell co-assembly in terms of the number of barcodes, by 33 vs. 112, it requires 7.1 Gbps sequencing which is more than that used by single cell co-assembly (3.5 Gbps). 

In all of our experiments, all distinct genomes were correctly detected. Therefore, our results exhibit ultimate sensitivity. However, in some experiments multiple cells with similar genomes were identified as distinct, which is not an issue for our problem because based on the results of {\tt Squeezambler 1.0}, those cells that are identified as with distinct genomes can be deeply sequenced and assembled afterwards. For the 62 cells, 6 distinct genomes collection, the number of detected distinct genomes was between 6 and 8. For the 97 cells, 5 distinct genomes collection, that number was between 5 and 11, and for the 112 cells, 7 distinct genomes collection, that was between 7 and 14. This specificity is reported in the sixth column of Table \ref{tab:results}. Note that the number of sequencing rounds (iterations) for single cell co-assembly is always 1, and for divide and conquer with $g = 2$, it is $\left\lceil \log_2 n \right\rceil$.

Due to the computational intensity of {\tt MDAsim 1.0}, {\tt HyDA 1.0}, and {\tt Squeezambler 1.0}, we report our results for only small examples  in order to provide a proof of concept. We also chose our parameters conservatively, and without optimization, so that we do not compromise accuracy. Moreover, our examples have in the order of 100 cells and 6 distinct genomes, whereas real world samples are much sparser as the number of cells may be in the order of billions and the number of distinct genomes at most in the order of thousands. Therefore, we expect the reduction in the total sequencing and maximum barcodes to be higher for real world applications than what we report in this paper.

\section{Conclusion}
We presented an adaptive divide and conquer algorithm for distilled sequencing and \emph{de novo} assembly of distinct genomes in a bacterial community, e.g. human gut microbiome. Samples derived from such communities are often sparse in the sense that the number of  distinct genomes is much less than the number of cells. Our algorithm exploits sparsity to decrease the amount of sequencing and the number of multiplexing barcodes needed for single cell sequencing and \emph{de novo} assembly. 

We implemented our algorithm in a tool which we call {\tt Squeezambler} and performed simulation experiments to demonstrate its power. Our results show that: i) the number of required barcodes with our divide and conquer algorithm is less than that required by  the na\"ive approach, and that ii) the amount of sequencing needed remains the same or decreases. Due to the computational intensity of the problem, only small examples with low sparsity were studied in this work. Real world samples are much sparser ($\sim1000$ species in $\sim10^{14}$ cells) than the examples here ($\sim5$ species in $\sim100$ cells). Also, the parameters used to run our tool were chosen conservatively and without optimization. Therefore, we expect the improvement of our algorithm to be higher than what we reported in this paper in real world situations. {\tt Squeezambler 1.0} identifies all distinct genomes in the sample which are candidates for different strains/species. Those cells that are identified as having distinct genomes need to be subsequently deeply sequenced and assembled  in order to obtain a more detailed assembly. 

\section*{Acknowledgement}

This work has been partially supported by the following grants: NIH RO1 DK089167, NIH STTR R42GM087013, and NSF DBI-0965741 (to S.D.). Any opinions, findings, and conclusions or recommendations expressed in this material are those of the authors and do not necessarily represent the views of any of the funding agencies.

%\paragraph{Funding\textcolon} 
%\bibliographystyle{achemnat}
\bibliographystyle{mynatbib}
\bibliography{pub,ref}
\end{document}